\pdfoutput=1

\documentclass[aps,showpacs,nofootinbib,superscriptaddress]{revtex4}
\usepackage{graphicx,subfigure}  
\usepackage{amsmath} 
\usepackage{amsfonts} 
\usepackage{amssymb} 
\usepackage{slashed} 

\newcommand{\sla}{\ensuremath{\slashed}}

\begin{document}
\title{Chiral dynamics in the $\vec{\gamma}p\rightarrow p\pi^0$ reaction}

\author{A. N. \surname{Hiller Blin}}
\affiliation{Departamento de F\'\i sica Te\'orica and IFIC, Centro Mixto
Universidad de Valencia-CSIC, Institutos de Investigaci\'on de
Paterna, E-46071 Valencia, Spain}
\author{T. \surname{Ledwig}}
\affiliation{Departamento de F\'\i sica Te\'orica and IFIC, Centro Mixto
Universidad de Valencia-CSIC, Institutos de Investigaci\'on de
Paterna, E-46071 Valencia, Spain}
\author{M. J.  \surname{Vicente Vacas}}
\affiliation{Departamento de F\'\i sica Te\'orica and IFIC, Centro Mixto
Universidad de Valencia-CSIC, Institutos de Investigaci\'on de
Paterna, E-46071 Valencia, Spain}

\today

\begin{abstract}
We investigate the neutral pion photoproduction on the proton near threshold in covariant chiral perturbation theory with the explicit inclusion of $\Delta$ degrees of freedom. This channel is specially sensitive to chiral dynamics and the advent of very precise data from the Mainz microtron has shown the limits of the convergence of the chiral series for both the heavy baryon and the covariant approaches. We show that the inclusion of the $\Delta$ resonance substantially improves the convergence leading to a good agreement with data for a wider range of energies.

\end{abstract}


\maketitle

\section{Introduction}
Neutral pion photoproduction on the proton at low energies is specially sensitive to chiral dynamics.
Considering  the range of energies from threshold to 500 MeV,
the total cross section appears to be clearly dominated by the magnetic dipole excitation of the $\Delta(1232)$\footnote{See, e.g., Fig. 8.1 of Ref.~\cite{Ericson:1988gk}}. Its role is more important here than for the  charged pions photoproduction, because of the smallness of the electric dipole contribution for the neutral pion channels.   Of course, approaching low energies, the relevance of the $\Delta$ resonance decreases fast and may become negligible as we get far from its mass and because of the $p$-wave nature of its contribution. 

Close to threshold, charged pion photoproduction has a relatively large cross section that  can be well described by just tree-level diagrams which lead to a substantial electric dipole moment. However, the situation is quite different for the neutral pion channels which present a much smaller cross section. Qualitatively, this is also well understood as the theoretical models produce a 
tiny $s$-wave amplitude, which actually vanishes in the chiral limit $(m_\pi\rightarrow 0)$. 
The smallness of the lowest order tree-level contributions offers a good opportunity for the study of higher order terms of the chiral Lagrangian and of loop effects. 
 In fact, one of the important successes of Chiral Perturbation Theory (ChPT) was the discovery in Refs.~\cite{Bernard:1991rt,Bernard:1992nc} of the importance of the loop contributions for the $\pi^0$ channels. This allowed  to solve the serious discrepancies between data 
\cite{Mazzucato:1986dz,Beck:1990da} and the Low Energy Theorems (LET) obtained by previous theoretical models \cite{DeBaenst:1971hp,Vainshtein:1972ih} based on current algebra and the partial conservation of the axial current \cite{Drechsel:1992pn,Bernard:2006gx}.
The model of Refs.~\cite{Bernard:1991rt,Bernard:1992nc} was further improved in Refs.
\cite{Bernard:1994gm,Bernard:1995cj} using a more systematic approach,  heavy-baryon ChPT (HBChPT), which allows for a proper power counting scheme. The neutral pion photoproduction off protons was analyzed  to fourth order in HBChPT in \cite{Bernard:2001gz} finding a good agreement with the data that were available at  the time.

However, the new and very precise data for the  $\vec\gamma+p\rightarrow \pi^0+p$ reaction obtained at the 
Mainz Microtron (MAMI) \cite{Hornidge:2012ca} have clearly shown the limits of this approach. 
In Ref. \cite{FernandezRamirez:2012nw}, it has been shown that fourth order HBChPT agrees well with data only up to around 20 MeV above threshold. 

 An alternative relativistic renormalization scheme of the baryons ChPT, the Extended On Mass Shell (EOMS) ChPT \cite{Gegelia:1999gf,Fuchs:2003qc} has been successfully applied to the study of several physical observables such as pion scattering, baryon magnetic moments and axial form factors, baryon masses among others
\cite{Fuchs:2003ir,Lehnhart:2004vi,Schindler:2006it,Schindler:2006ha,Geng:2008mf,Geng:2009ik,MartinCamalich:2010fp,Alarcon:2011zs,Chen:2012nx,
Ledwig:2014rfa}. The EOMS approach is covariant, satisfies analyticity constraints lost in the HB formulation and usually converges relatively faster. Surprisingly, a fourth order EOMS calculation of the $\vec\gamma+p\rightarrow \pi^0+p$  process described the experimental data even slightly  worse than the HB one~\cite{Hilt:2013uf,Hornidge:2012ca}.  

A possible reason for the poor agreement could be due to the importance of the $\Delta$ resonance, not included in the aforementioned calculations as an explicit degree of freedom. Here, it could be more visible than for other channels due to the smallness of the nucleonic contributions of the lowest orders. This was already pointed out by Hemmert et al. in Ref.~\cite{Hemmert:1996xg}. Actually, they obtained a moderate effect for the electric dipole amplitude at threshold in their HB approach. This result was further explored in Ref.~\cite{Bernard:2001gz}, also in a HBChPT static calculation, finding a sizable cancellation of the $\Delta$ contributions by fourth order loop effects.
 However, it could be expected that, in a dynamical calculation (with the full $\Delta$  propagator), the effects could grow very fast as a function of the photon energy as the invariant mass of the system at threshold is close to the resonance mass ($M_\Delta-E_{CM}\sim m_\pi$).
Of course, the $\Delta$ effects could be accounted for by a change in the Low Energy Constants (LECs) and by higher order terms. However, if the $\Delta$ resonance plays an important role, its inclusion could lead to a faster convergence and more natural values of the LECs.  The possible relevance of the $\Delta$ mechanisms for this process was also signaled in Refs.~\cite{Hornidge:2012ca,FernandezRamirez:2012nw}.   

Our purpose in this work is to explore the influence of the $\Delta$ mediated mechanisms in the photoproduction of $\pi^0$ off protons. We will calculate the process in the purely nucleonic EOMS ChPT scheme up to order $p^3$ and will add 
the $\Delta$ resonance contribution at tree level. We will compare our results with the precise data on the near threshold angular cross sections and photon asymmetries from Ref.~\cite{Hornidge:2012ca} and study the range of validity of our expansion.

\section{Theoretical Model}

For an $O(p^3)$ calculation, the relevant terms of the chiral Lagrangian, including only pions, nucleons and photons as degrees of freedom are shown below with the superscript indicating the chiral order. We follow the naming conventions for the LECs from \cite{Fettes:2000gb}. At first order we have  
\begin{equation}
\mathcal{L}_N^{(1)}=\bar{\Psi}\left(\mathrm{i}\sla{\mathrm{D}}-m+\frac{g_A}{2}\sla u\gamma_5\right)\Psi,
\end{equation}
where $\Psi$ is the nucleon doublet with mass $m$ and $\mathrm{D}_\mu=\left(\partial_\mu+\Gamma_\mu\right)$ is the covariant derivative given by
\begin{equation*}
\Gamma_\mu=\frac12\left[u^\dagger(\partial_\mu-\mathrm{i}r_\mu)u
+u(\partial_\mu-\mathrm{i}l_\mu)u^\dagger\right]. 
\end{equation*}
The meson fields appear through 
\begin{equation*}
u=\exp\left(\frac{\mathrm{i}\phi}{2F}\right), \quad
 \phi=\left(\begin{array}{cc}
\pi^0&\sqrt2\pi^+\\
\sqrt2\pi^-&-\pi^0
\end{array}\right),
\end{equation*}
with $F$ the pion decay constant, and also in 
$
u_\mu=\mathrm{i}\left[u^\dagger(\partial_\mu-\mathrm{i}r_\mu)u
-u(\partial_\mu-\mathrm{i}l_\mu)u^\dagger\right].
$
The photon field $\mathcal{A}_\mu$ couples through 
\begin{equation*}
r_\mu=l_\mu=\frac e2\mathcal{A}_\mu(\mathbb{I}_2+\tau_3),
\end{equation*}
where $\tau_3$ is the Pauli matrix and $e$ is the (negative) electron charge. At second order, there are only two relevant terms 
\begin{equation}
\mathcal{L}_N^{(2)}=\frac{1}{8m}\bar{\Psi}\left(
c_6 f_{\mu\nu}^+ + c_7\text{Tr}\left[f_{\mu\nu}^+\right]
\right)\sigma^{\mu\nu}\Psi+\dots,
\end{equation}
where $ f_{\mu\nu}^+=uf_{\mu\nu}^Lu^\dagger+u^\dagger f_{\mu\nu}^Ru$ and for our case
$
f_{\mu\nu}^{R}=f_{\mu\nu}^{L}=\partial_\mu r_\nu - \partial_\nu r_\mu -\mathrm{i}\left[r_\mu,r_\nu\right].
$
Finally, at third order we have 
\begin{align}
 \mathcal{L}_N^{(3)}= &d_8\frac{\mathrm{i}}{2m}\left\{
\bar\Psi\varepsilon^{\mu\nu\alpha\beta}\text{Tr}\left[\tilde f_{\mu\nu}^+u_\alpha\right]
\mathrm{D}_\beta\Psi
\right\}+\text{h.c.}\\
 +&d_9\frac{\mathrm{i}}{2m}\left\{
\bar\Psi\varepsilon^{\mu\nu\alpha\beta}\text{Tr}\left[f_{\mu\nu}^+
\right]u_\alpha
\mathrm{D}_\beta\Psi
\right\}+\text{h.c.}\nonumber\\
 +&d_{16}\frac{1}{2}\left\{
\bar\Psi\gamma^{\mu}\gamma_5\text{Tr}\left[\chi_+\right]u_\mu\Psi
\right\}\nonumber\\
 +&d_{18}\frac{\mathrm{i}}{2}\left\{
\bar\Psi\gamma^{\mu}\gamma_5[\mathrm{D}_\mu,\chi_-]\Psi
\right\}+\dots,\nonumber
\end{align}
where
$
 \tilde f_{\mu\nu}^+ = f_{\mu\nu}^+-\frac12\text{Tr}\left[f_{\mu\nu}^+\right],$ 
$
\chi_\pm=u^\dagger\chi u^\dagger\pm u\chi^\dagger u$. We will work in the isospin limit as it was done in Ref. \cite{Hilt:2013uf}, hence $\chi=m_\pi^2$, the pion mass 
squared\footnote{This means that we cannot study the cusp effects appearing at the opening of the charged pion channels.
Formally, the error introduced by the use of a single value for the pion mass in the calculation of the loops is of higher order.}.
We also need the purely mesonic term
\begin{align}
\mathcal{L}_\pi^{(2)}=\frac{F^2}{4}\text{Tr}{\left[
\mathrm{D}_\mu U(\mathrm{D}^\mu U)^\dagger+\chi U^\dagger +U \chi^\dagger
\right]},
\end{align}
where $U=u^2$ and whose covariant derivative acts as
$
\mathrm{D}_\mu U=\partial_\mu U - \mathrm{i}r_\mu U + \mathrm{i}U l_\mu.
$

At $O(p^3)$, there is a large number of contributions to the pion photoproduction process, including both tree-level diagrams and loops. A full set of the loop diagrams can be found, e.g., in Ref.~\cite{Bernard:1992nc}. In the next figures, we show the relevant diagrams for our specific channel (real photons, neutral pions). We also omit the crossed ones. 

\begin{figure}
\subfigure[]{
\label{fO1a}
\includegraphics[width=0.3\textwidth]{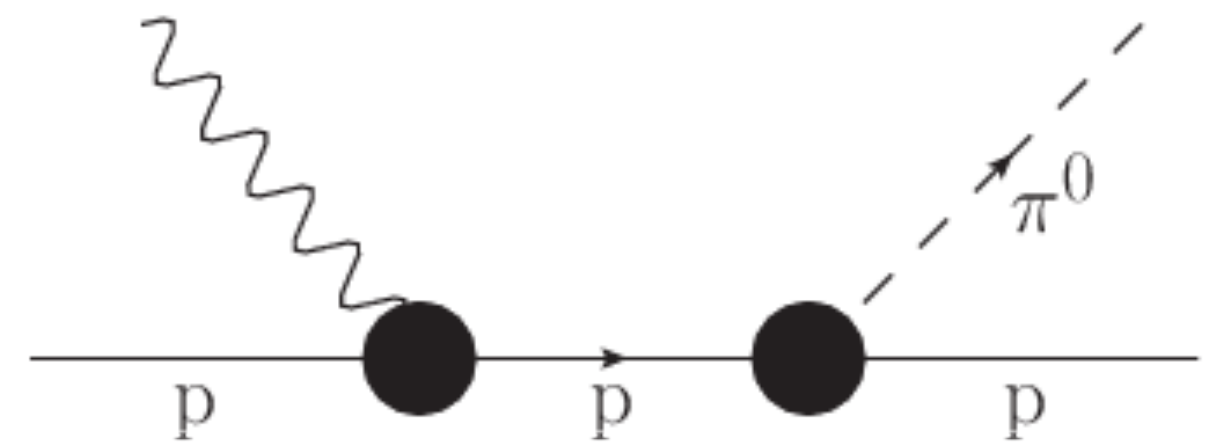}}
\subfigure[]{
\label{fO1c}
\includegraphics[width=0.2\textwidth]{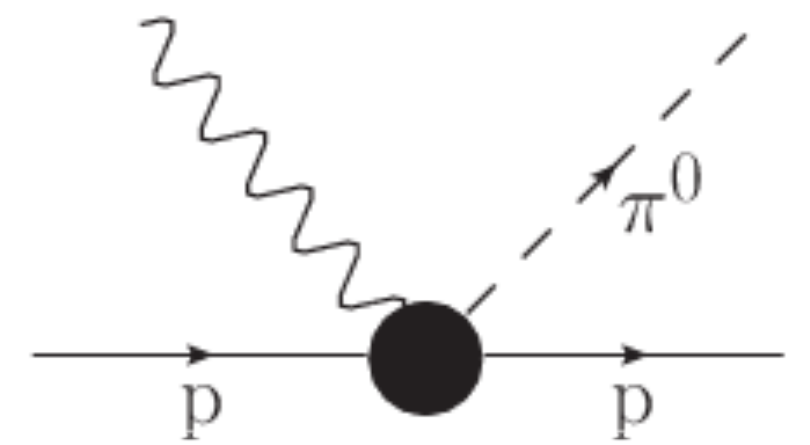}}
\caption{Tree diagrams for  $\pi^0$ photoproduction off protons. The black dots represent vertices of chiral order 1 to 3. Diagram (b) starts at order 3.}
\label{fig:1}
\end{figure}

In Fig.~\ref{fig:1}, we show the tree-level diagrams. Both, the $\gamma NN$ and the $\pi NN$ vertices contain pieces of chiral order running  from one to three. However, the contact $\gamma\pi^0 pp$ term starts at third order.

\begin{figure}
\begin{center}
\subfigure{
\label{fO3LMa}
\includegraphics[width=0.3\textwidth]{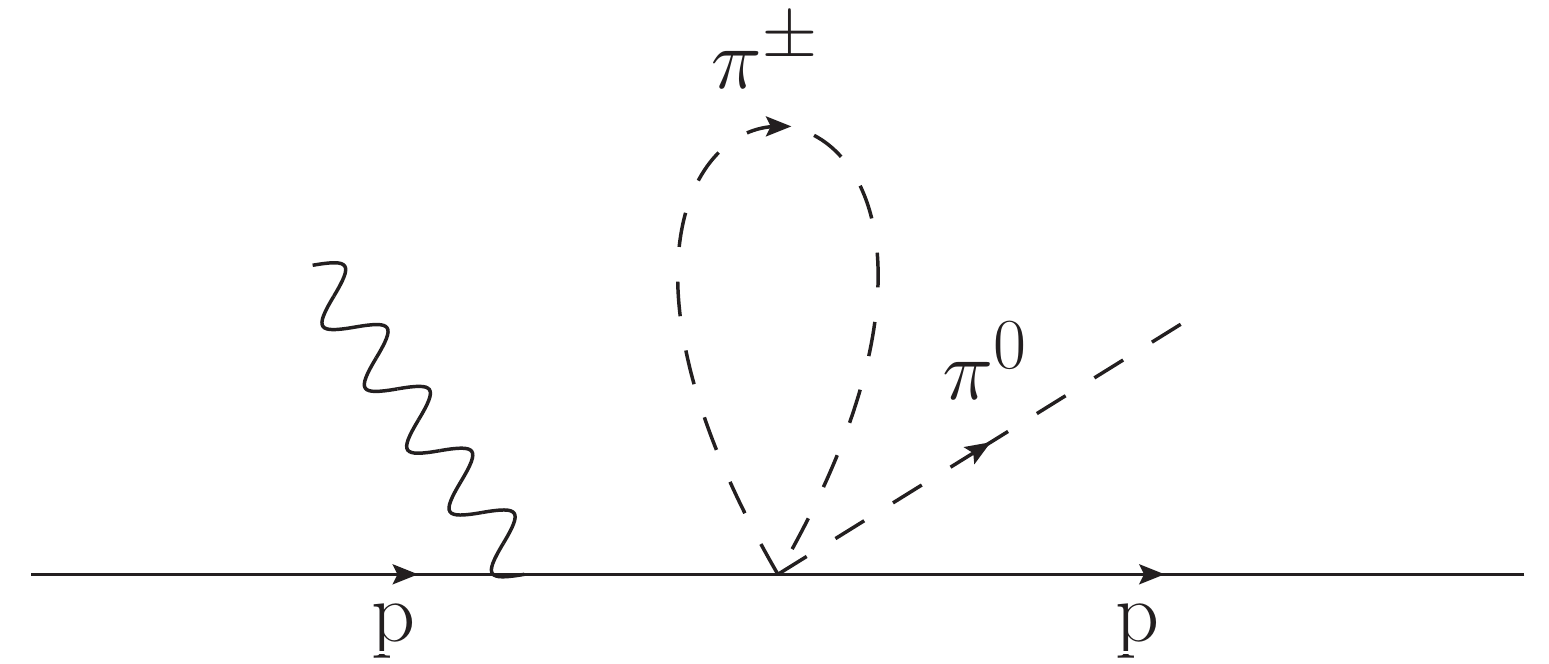}}
\subfigure{
\label{fO3LMc}
\includegraphics[width=0.3\textwidth]{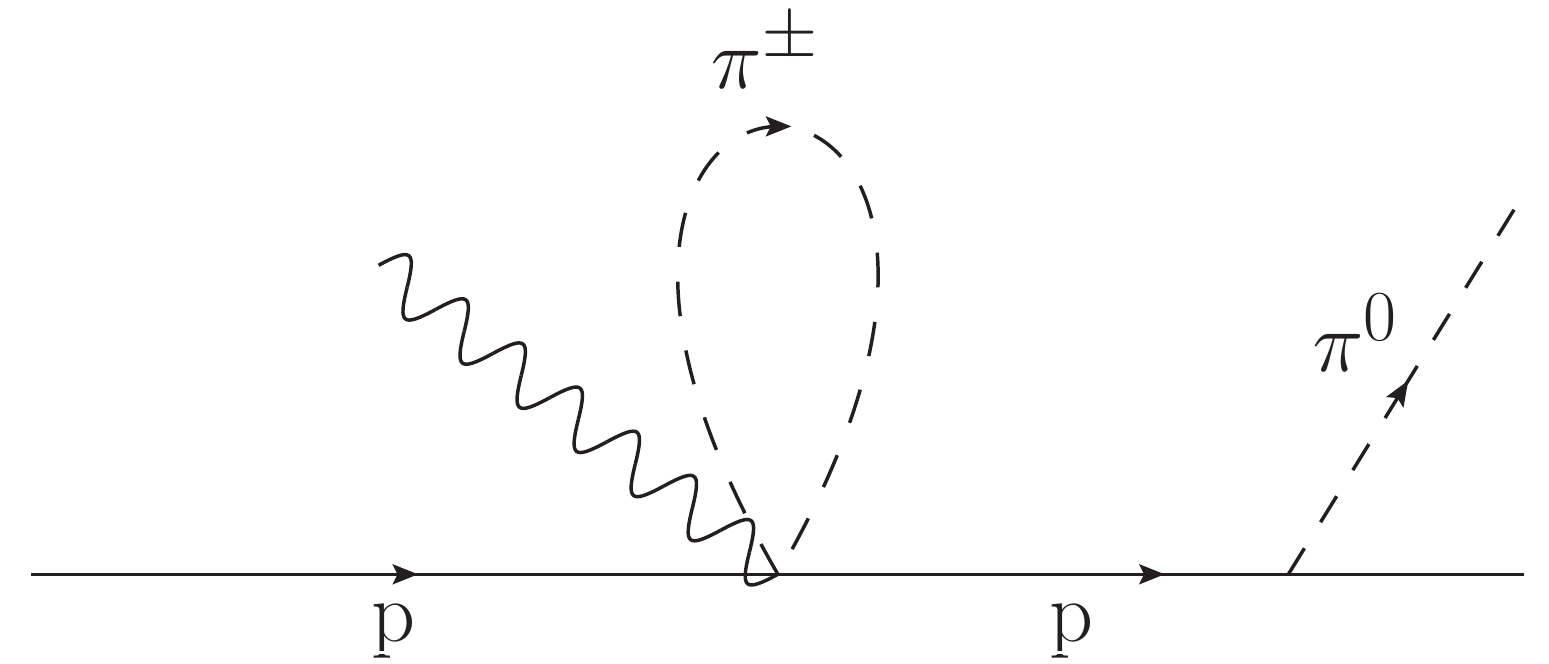}}
\subfigure{
\label{fO3LMe}
\includegraphics[width=0.3\textwidth]{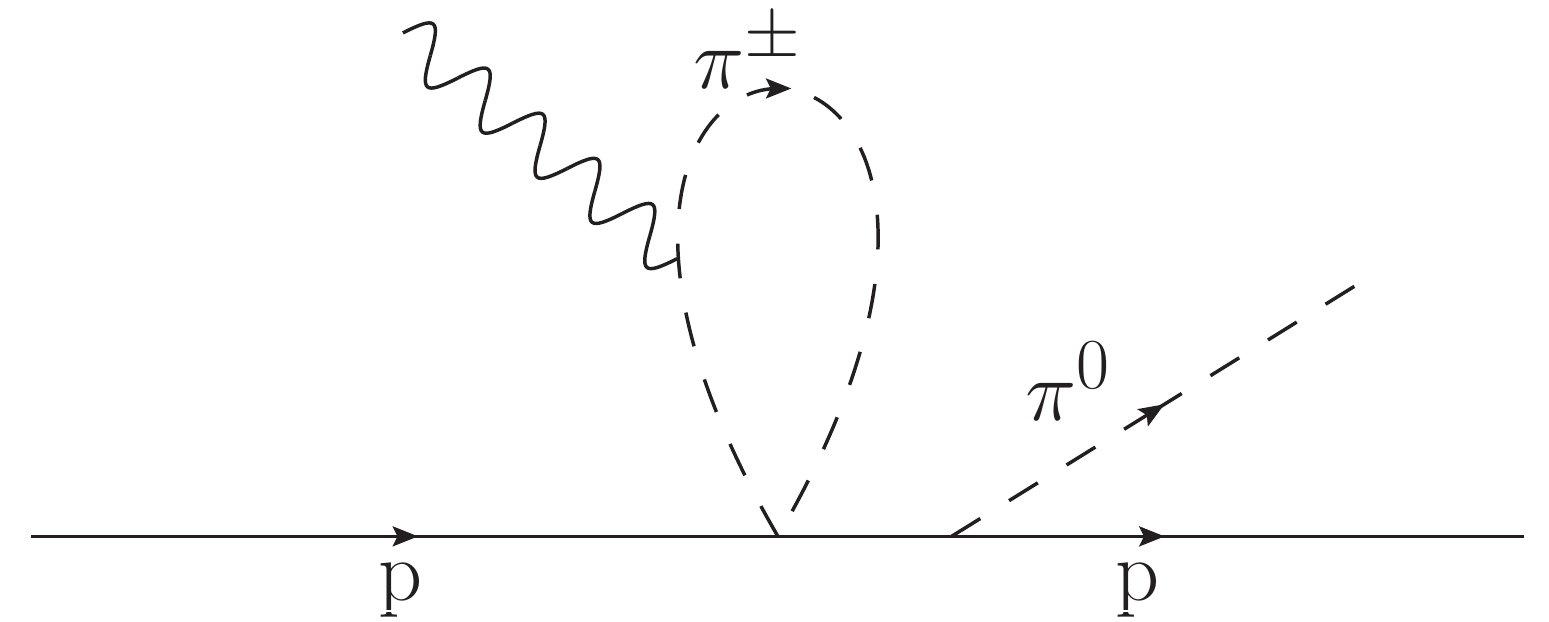}}
\end{center}
\caption{Diagrams with purely mesonic loops.}\label{fig:2}
\end{figure}

\begin{figure}
\begin{center}
\subfigure{
\label{fO3LB1a}
\includegraphics[width=0.3\textwidth]{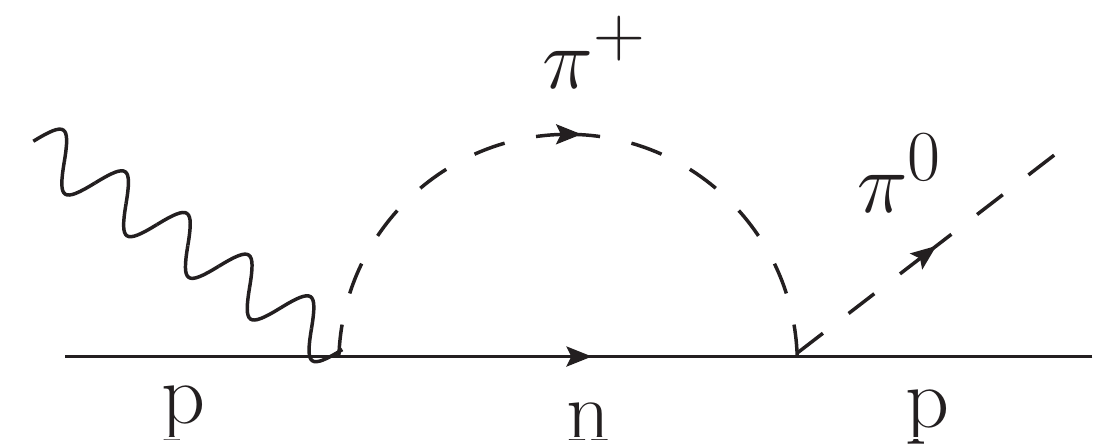}}
\subfigure{
\label{fO3LB1b}
\includegraphics[width=0.3\textwidth]{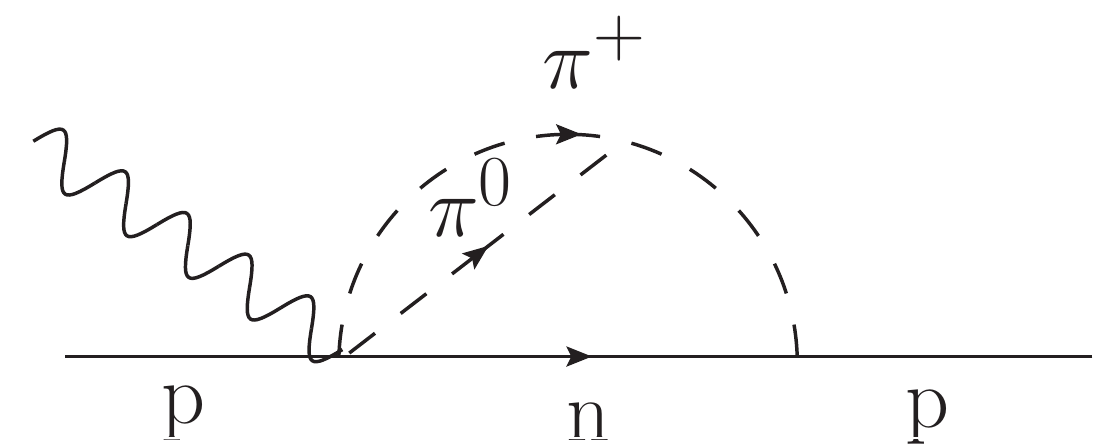}}
\subfigure{
\label{fO3LB1e}
\includegraphics[width=0.3\textwidth]{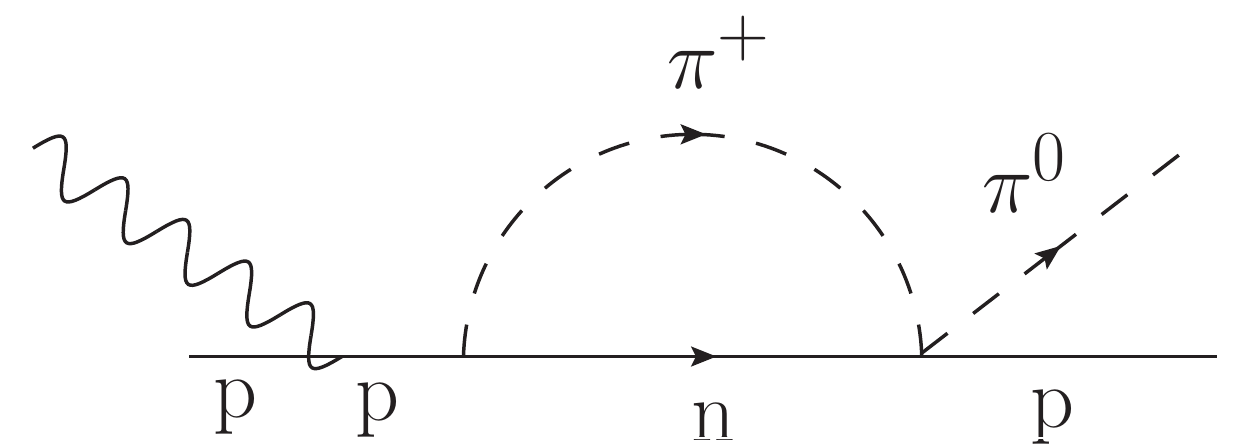}}\\
\subfigure{
\label{fO3LB1i}
\includegraphics[width=0.3\textwidth]{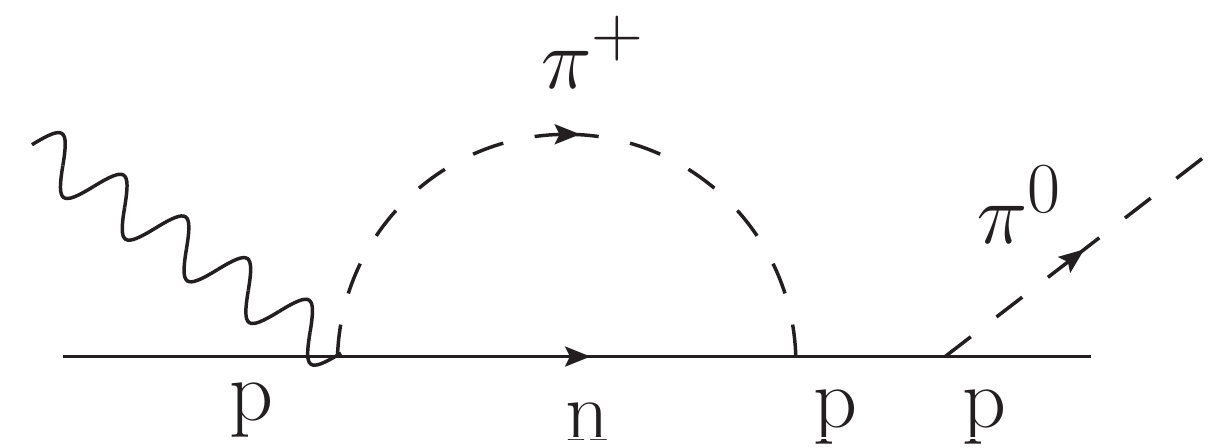}}
\subfigure{
\label{fO3LB1m}
\includegraphics[width=0.3\textwidth]{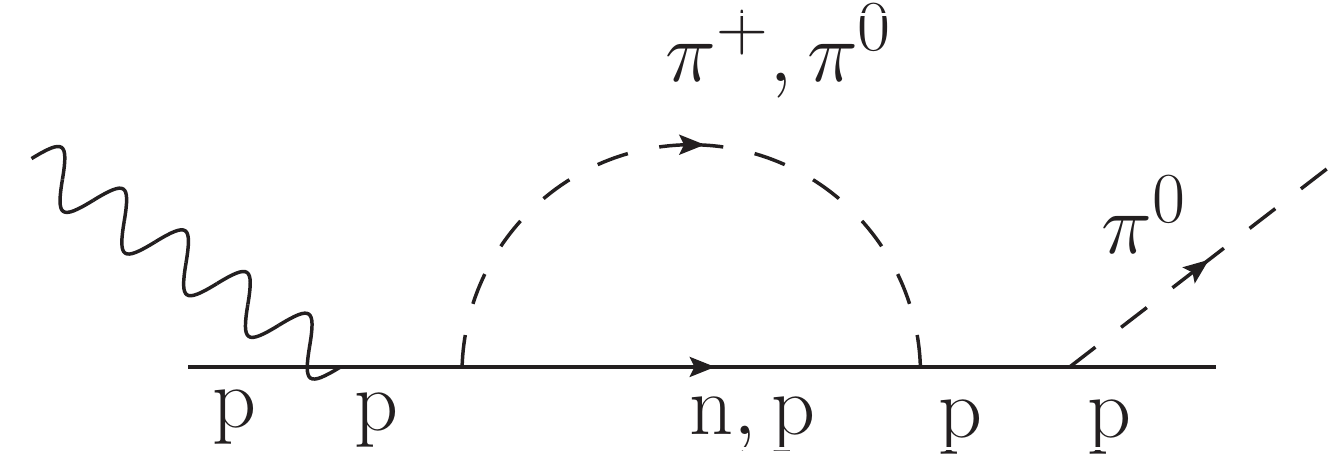}}
\end{center}
\caption{Loop diagrams with one meson and one baryon in the loop.}\label{fig:3}
\end{figure}

\begin{figure}
\begin{center}
\subfigure{
\label{fO3LBM3a}
\includegraphics[width=0.3\textwidth]{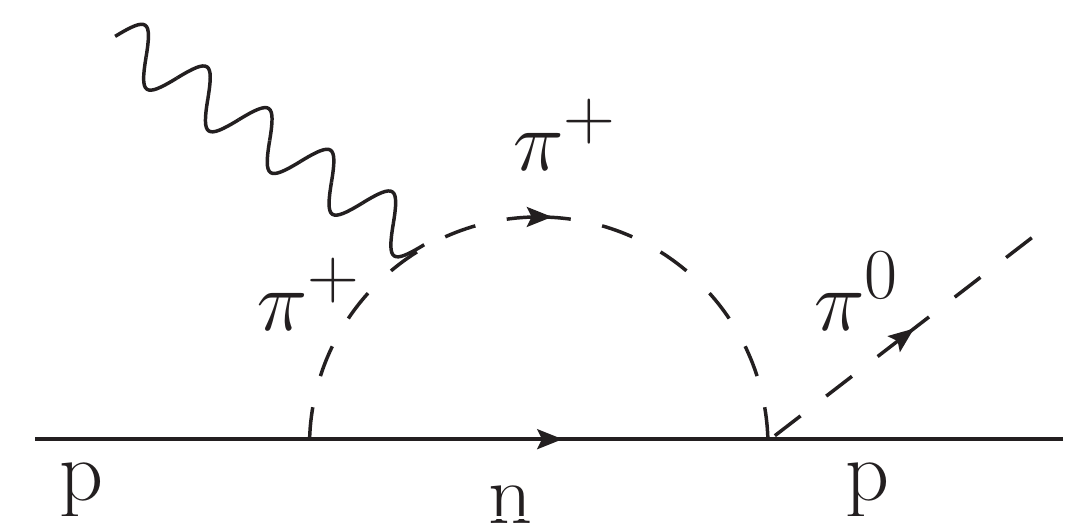}}
\subfigure{
\label{fO3LBM3c}
\includegraphics[width=0.3\textwidth]{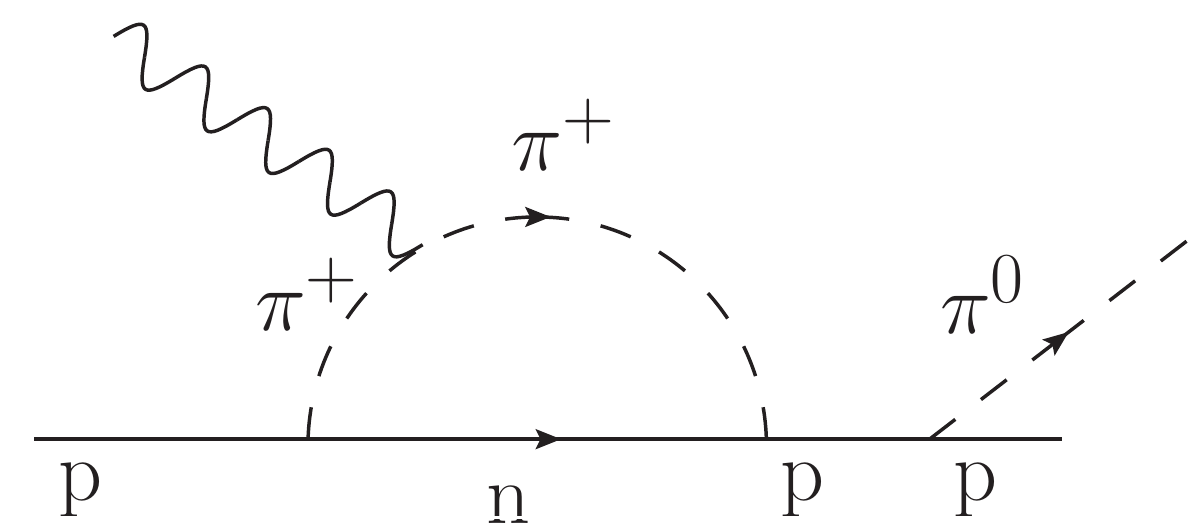}}
\subfigure{
\label{fO3LBM3e}
\includegraphics[width=0.3\textwidth]{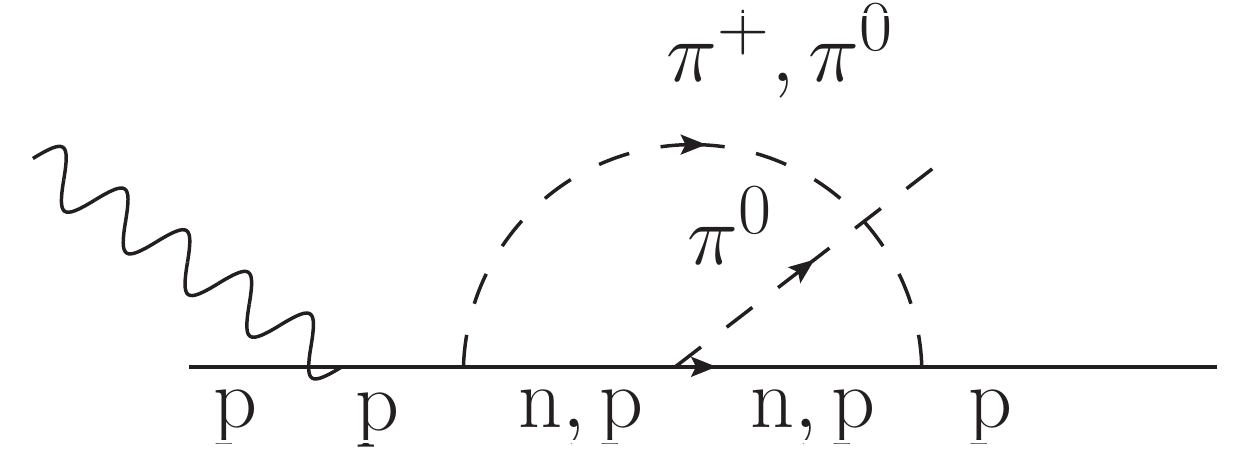}}\\
\subfigure{
\label{fO3LBM3g}
\includegraphics[width=0.3\textwidth]{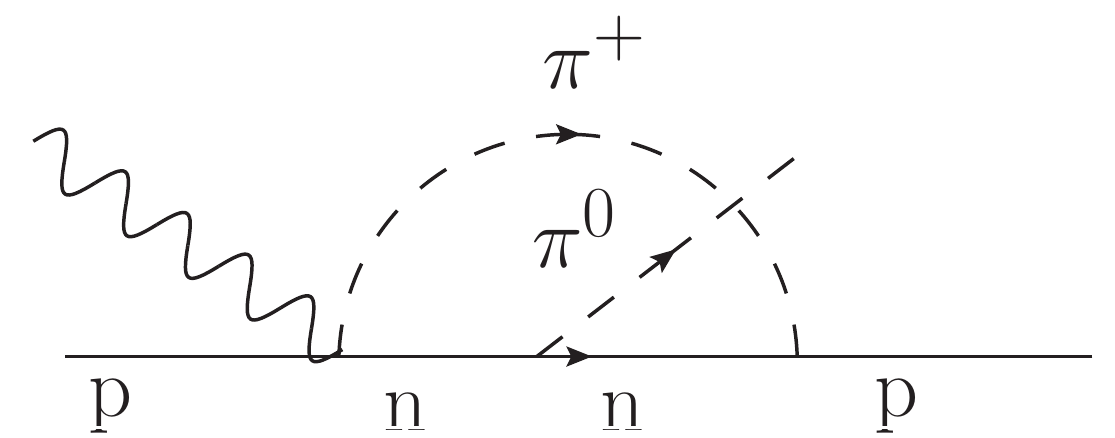}}
\subfigure{
\label{fO3LBM3i}
\includegraphics[width=0.3\textwidth]{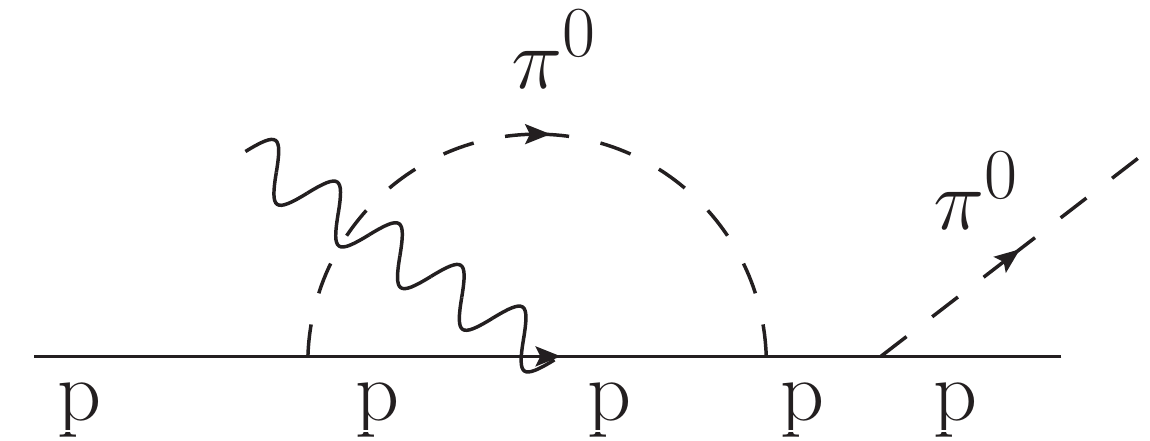}}
\end{center}
\caption{Loop diagrams with three hadron propagators in the loop.}\label{fig:4}
\end{figure}

\begin{figure}
\begin{center}
\subfigure{
\label{fO3LBM4a}
\includegraphics[width=0.3\textwidth]{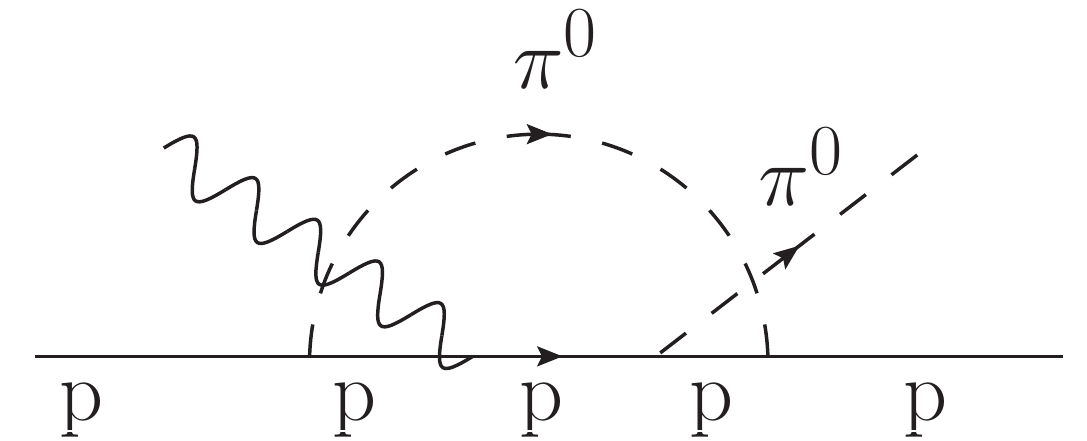}}
\subfigure{
\label{fO3LBM4b}
\includegraphics[width=0.3\textwidth]{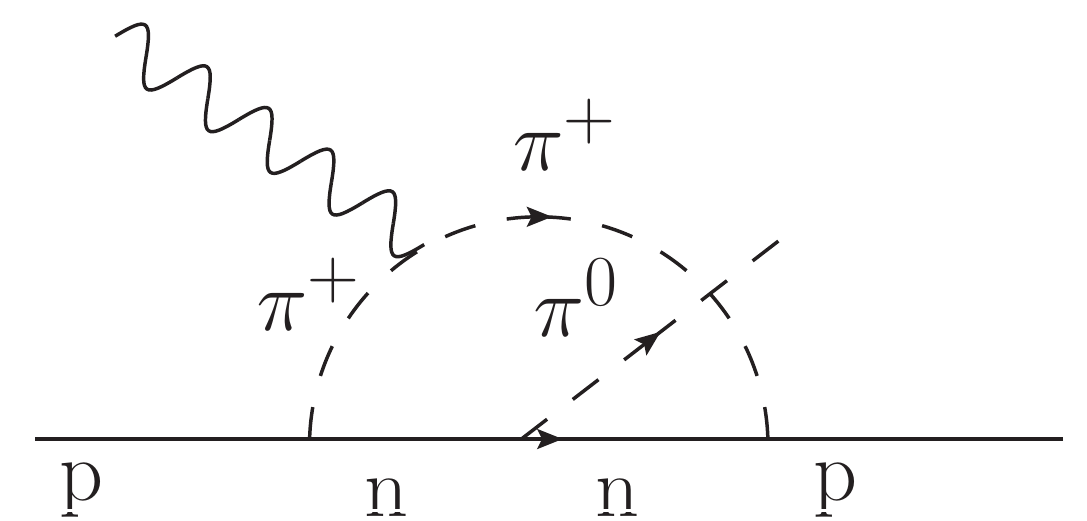}}
\end{center}
\caption{Loop diagrams with four hadron propagators in the loop.}\label{fig:5}
\end{figure}

The loop terms, contributing up to $O(p^3)$,  are depicted in Figs. \ref{fig:2}, \ref{fig:3}, \ref{fig:4} and \ref{fig:5}. 
The loop diagrams have been evaluated applying the EOMS renormalization scheme. First, 
we have removed the infinities using the modified minimal subtraction $(\widetilde{MS})$ scheme \cite{Scherer:2012xha}. Then, after making an expansion of the amplitudes\footnote{We have chosen for the expansion the three small parameters $m_\pi$, $\nu=(s-u)/(4 m)$ with $s$ and $u$ the Mandelstam variables of order one and  the Mandelstam variable $t$ of order 2 as in Ref.\cite{Alarcon:2012kn}.}
we have removed the power counting breaking terms (those with a chiral order lower than the nominal order of the loop). 
Obviously, the diagrams from Fig.~\ref{fig:2}, which contain exclusively mesonic loops, do not break the power counting.

Furthermore, we have to consider the wave function renormalization of the external legs. In our calculation we only include it  at $O(p^2)$ on the external proton legs of the tree diagrams of $O(p^1)$ as all other corrections are at least $O(p^4)$. This amounts to multiplying the amplitude obtained for those terms by a factor 
\begin{equation}
    Z_p= 1+\frac{3 g_A^2 m_\pi^2}{32\pi^2F^2}\left(3 \log\left(\frac{m}{m_\pi}\right)-2\right).
\end{equation} 

Finally, we should mention that, apart from $g_A$, at $O(p^3)$, the $\gamma+p\rightarrow p+\pi^0$ scattering amplitude depends only on some specific combinations of the LECs: $\tilde{d}_{89}=d_8+d_9$, $\tilde{d}_{168}=2d_{16}-d_{18}$ and 
$\tilde{c}_{67}=c_6+c_7$.

The electromagnetic excitation of the $\Delta(1232)$ has been much investigated since the late fifties~\cite{Chew:1957tf,Adler:1968tw}. Most of the work has dealt with energies around the resonance region, where the $\Delta$ usually plays a dominant role.  In the last years, we could mention the
 review of Ref.~\cite{Pascalutsa:2006up} and, e.g., some  works on pion electro- and photoproduction \cite{Pascalutsa:2004pk,Pascalutsa:2005vq,FernandezRamirez:2005iv}, or Compton scattering evaluated  in covariant ChPT \cite{Lensky:2009uv}. There are also some recent advances incorporating the $\Delta$ as a dynamic degree of freedom in the analysis of $\pi N$ scattering \cite{Alarcon:2012kn} in the same EOMS approach that we use here. 

For the neutral pion photoproduction close to threshold, the $\Delta$ isobar effects have been calculated in HBChPT \cite{Hemmert:1996xg,Bernard:2001gz}, in a static approach, obtaining only moderate effects.

\begin{figure}
\begin{center}
\label{fDd}
\includegraphics[width=0.45\textwidth]{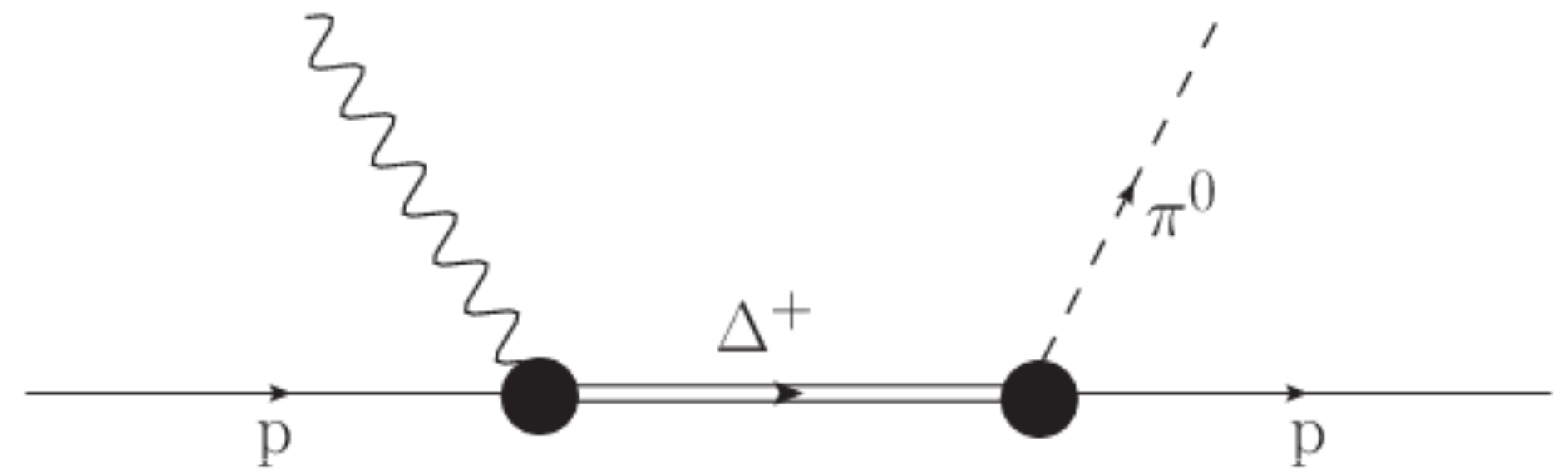}
\end{center}
\caption{$\Delta$ tree diagram for $\pi^0$ photoproduction off protons.}\label{fig:6}
\end{figure}

Here, we will consider the tree-level $\Delta$ resonance diagrams, which include the direct one from Fig.~\ref{fig:6} and the crossed one, in a dynamic fashion maintaining the energy dependence of the resonance propagator. To describe the $\Delta$ interactions  we use consistent Lagrangians which ensure the decoupling of the spurious spin-1/2 components of the Rarita-Schwinger field. The relevant pieces are
\begin{align}
\mathcal{L}^ {(1)}_{\Delta\pi N}=&\frac{\mathrm{i}h_A}{2FM_\Delta}\bar{\Psi}T^a\gamma^{\mu\nu\lambda}(\partial_\mu\Delta_\nu)(\partial_\lambda\pi^a) + \text{h.c.}\\
\mathcal{L}^ {(2)}_{\Delta\pi N}=&\frac{h_1}{2FM_\Delta^2}\bar{\Psi}T^a\gamma^{\mu\nu\lambda}(\partial_\lambda\slashed\partial\pi^a)(\partial_\mu\Delta_\nu) + \text{h.c.}\\
  \mathcal{L}^ {(2)}_{\Delta\gamma N}=&\frac{3\mathrm{i}eg_M}{2m(m+M_\Delta)}\bar{\Psi}T^3(\partial_\mu\Delta_\nu)\tilde{F}^ {\mu\nu} + \text{h.c.}\\
  \mathcal{L}^ {(3)}_{\Delta\gamma N}=&-\frac{3eg_E}{2m(m+M_\Delta)}\bar{\Psi}T^3\gamma_5(\partial_\mu\Delta_\nu)F^ {\mu\nu} + \text{h.c.},
\end{align}
where $F^ {\mu\nu}$   and $\tilde{F}^ {\mu\nu}$ are the electromagnetic field and its dual. There are two couplings for the pion ($h_A$, $h_1$) and two for the photon, the magnetic piece ($g_M$) of chiral order two and the electric piece ($g_E$) of order three. At third order, the Lagrangian contains an additional $\gamma N\Delta $ Coulomb coupling  which vanishes for real photons. The conventions and definitions for the isospin operators $T$ can be found in Ref.~\cite{Pascalutsa:2002pi}. Actually, we neglect the  $h_1$ piece in our calculation for simplicity and because its value has been found to be consistent with zero~\cite{Pascalutsa:2006up}. For the other constants, we take $h_A=2.85$, $g_M=2.97$ and $g_E=-1.0$ \cite{Pascalutsa:2005vq}. The value for $h_A$ can be directly obtained from the $\Delta$ width, and $g_M$ and $g_E$ were obtained  fitting  pion electromagnetic production at energies around the resonance peak.

In the standard chiral counting scheme for diagrams without $\Delta$ resonances, the order $D$ of a diagram with $L$ loops, 
$V_{k}$ vertices from $\mathcal{L}^{(k)}$, $N_\pi$ pionic propagators and $N_N$ nucleonic propagators is given by
\begin{equation}
D=4L+\sum{kV_{k}}-2N_\pi-N_N.
\end{equation}
Another small parameter, $\delta= M_\Delta - m\sim 300\text{MeV}$, appears when we introduce the $\Delta$ resonance and several prescriptions have been used in the literature to establish an appropriate power counting scheme for this case \cite{Hemmert:1996xg,Pascalutsa:2002pi}. Here, we follow the ``$\delta$ counting'' scheme. In our low-energy range, very close to the pion production threshold, we count $\delta^2$ as being of $\mathcal{O}(p)$, following Ref.~\cite{Lensky:2009uv}. Hence, one obtains the rule
\begin{align}
D=4L+\sum{kV_{k}}-2N_\pi-N_N-\frac{1}{2}N_\Delta,
\end{align}
where now $N_\Delta$ is the number of $\Delta$ propagators. Thus, we have taken into account all the amplitudes up to order $D=3$ according to this counting rule, as well as the $\Delta$ tree-level contribution  proportional to $g_E$ of order $D=3.5$. The effect of this latter piece is negligible. The mechanisms including $\Delta$ loops would start contributing at $D=3.5$. 

\section{Results}
We compare our model with the full set of data  of 
Refs.~\cite{Hornidge:2012ca,Hilt:2013uf,Hornidge:pri} on the angular cross section and $\Sigma$, the linearly polarized photon asymmetry
\begin{equation}
\Sigma=\frac{d\sigma_\perp-d\sigma_\parallel}{d\sigma_\perp+d\sigma_\parallel},
\end{equation}
with $d\sigma_\perp$ and $d\sigma_\parallel$ the angular cross sections for photon polarization perpendicular and parallel to the reaction plane with the pion and the outgoing proton.

The analysis of these data has shown that  HBChPT agrees well  only up to approximately 20 MeV above threshold
\cite{FernandezRamirez:2012nw} and covariant ChPT does it even worse \cite{Hornidge:2012ca}. This result was nicely shown by studying the $\chi^2$ per degree of freedom  as a function of the maximum photon energy of the data included in the fit. 

\begin{figure}
\begin{center}
\includegraphics[scale=.56,angle=0]{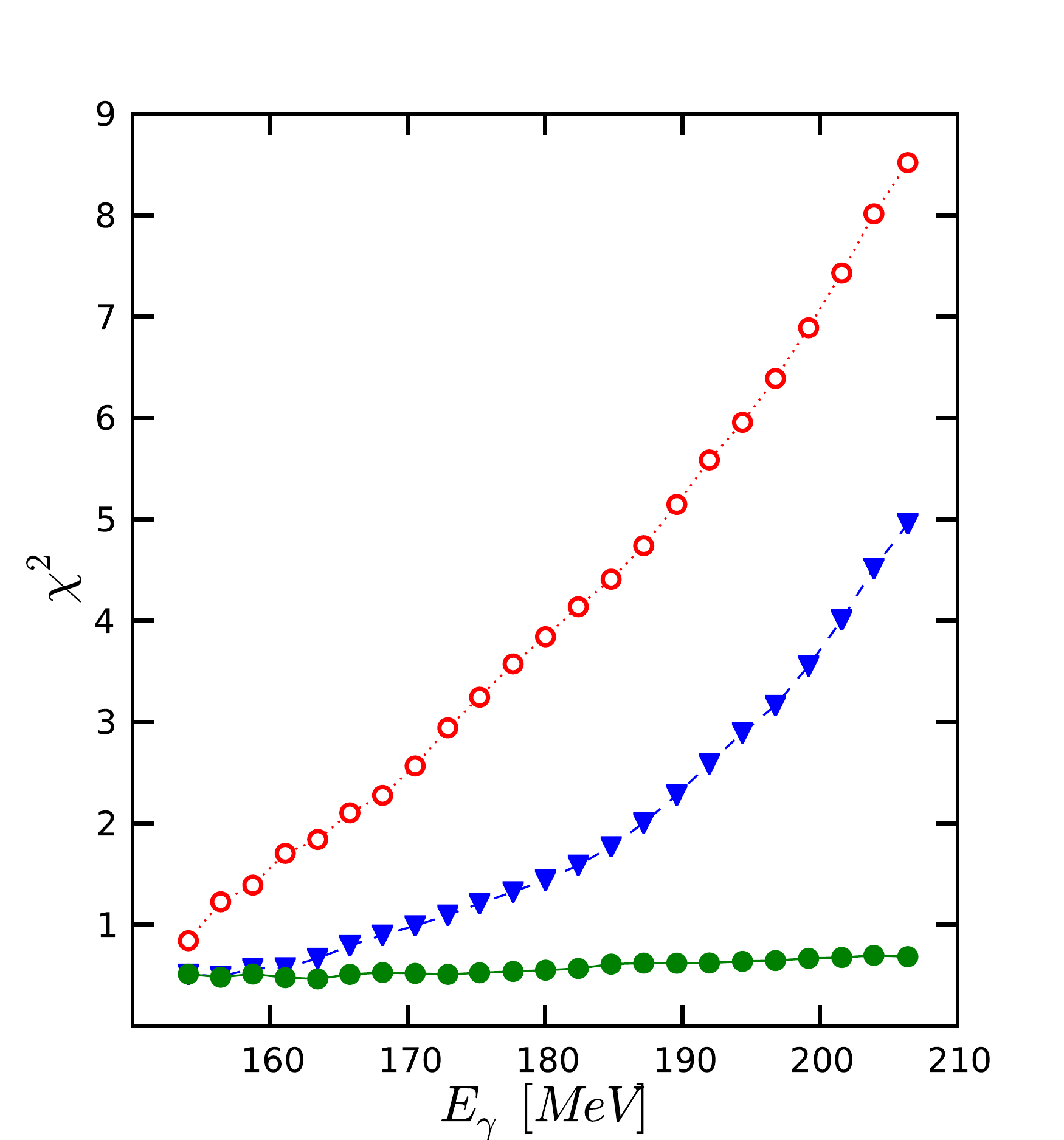} 
\end{center}
\caption{$\chi^2$ per degree of freedom as a function of the maximum photon energy of the data included in the fit. 
Solid circles: full model, triangles: model without the $\Delta$ resonance, empty circles:  only tree-level nucleonic contributions (without $\Delta$ and loops). Lines to guide the eye.}
\label{fig:7}
\end{figure}

In Fig.~\ref{fig:7}, we show our results for this magnitude. We have fixed $m$ and $m_\pi$ to their physical values
for proton and neutral pion, $F=92.4$ MeV and the $\Delta$ couplings as given in the previous section.
We prefer to fix $g_M$ and $g_E$, even when the latter one is poorly known. In principle, these values could also be fitted, but a more comprehensive analysis including other charge channels and a wider range of energies, where the $\Delta$ mechanisms could be dominant, would be better suited for that purpose. As we will see below, at low  energies and for our channel, the size of the $\Delta$ contribution is relatively small, even when it is essential to get a good agreement with data. 
The rest of the LECs, $g_A$, $\tilde{c}_{67}$, $\tilde{d}_{89}$ and $\tilde{d}_{168}$ have been taken as free parameters. 

 We start our fit at energies above the charged pion threshold to avoid the cusp effects\footnote{Including the few missing points does not  appreciably modify the results.}.
 The loop contributions improve the agreement at the threshold region, showing the remarkable effect already found in Ref.~\cite{Bernard:1991rt}. Still,
 the $\chi^2$ of our model with tree-level and loop diagrams up to $O(p^3)$ and just nucleons grows quickly as a function of the photon energy and qualitatively reproduces the previous results of Refs. \cite{Hornidge:2012ca,Hilt:2013uf, FernandezRamirez:2012nw}. This is only to be expected, as \cite{Hilt:2013uf}, which corresponds to a higher order calculation in the same EOMS covariant ChPT approach used here, with further free parameters, could not reach a good agreement over the whole energy range.

The situation changes drastically as soon as the $\Delta$ mechanisms are included, even when it does not imply any new free parameter. As an additional check, we also let free $g_M$, the dominant magnetic $N\Delta$ coupling, and we obtain for the best fit $g_M=3.1$, which is very close to the value taken from the literature. Although this could suggest that there is enough information on the current data to fix $\Delta$  LECs, a more general study would be convenient, because higher order terms might modify this result.
The values of the LECs for the different cases studied can be seen in Tab.~\ref{tab:1}. In particular, we could fix $g_A$ to its physical value without altering the quality of the fit, as the effects of the modification are absorbed by changes in the other LECs.

\begin{table}
\begin{center}
\begin{tabular}{|l|c|c|c|c|c|}
\hline 
  & $g_A$ & $\tilde{c}_{67}$ & $\tilde{d}_{89}$ [GeV$^{-2}$] & $\tilde{d}_{168} $ [GeV$^{-2}$]& $\chi^2$/d.o.f. \\ 
\hline 
No $\Delta$ & 1.46 & 2.86 & 4.20 & -15.1 & 4.96 \\ 
\hline 
Full model & {\bf 1.27} & 2.33 & 1.46 & -12.1 & 0.69 \\ 
\hline 
Full model & 1.24 & 2.36 & 1.46 & -11.1 &  0.68\\ 
\hline 
\end{tabular} 
\end{center}
\caption{LEC values in different versions of the model. Fixed values appear in boldface.}\label{tab:1}
\end{table}

\begin{figure}
\begin{center}
\subfigure{
\label{f8a}
\includegraphics[width=1.08\textwidth]{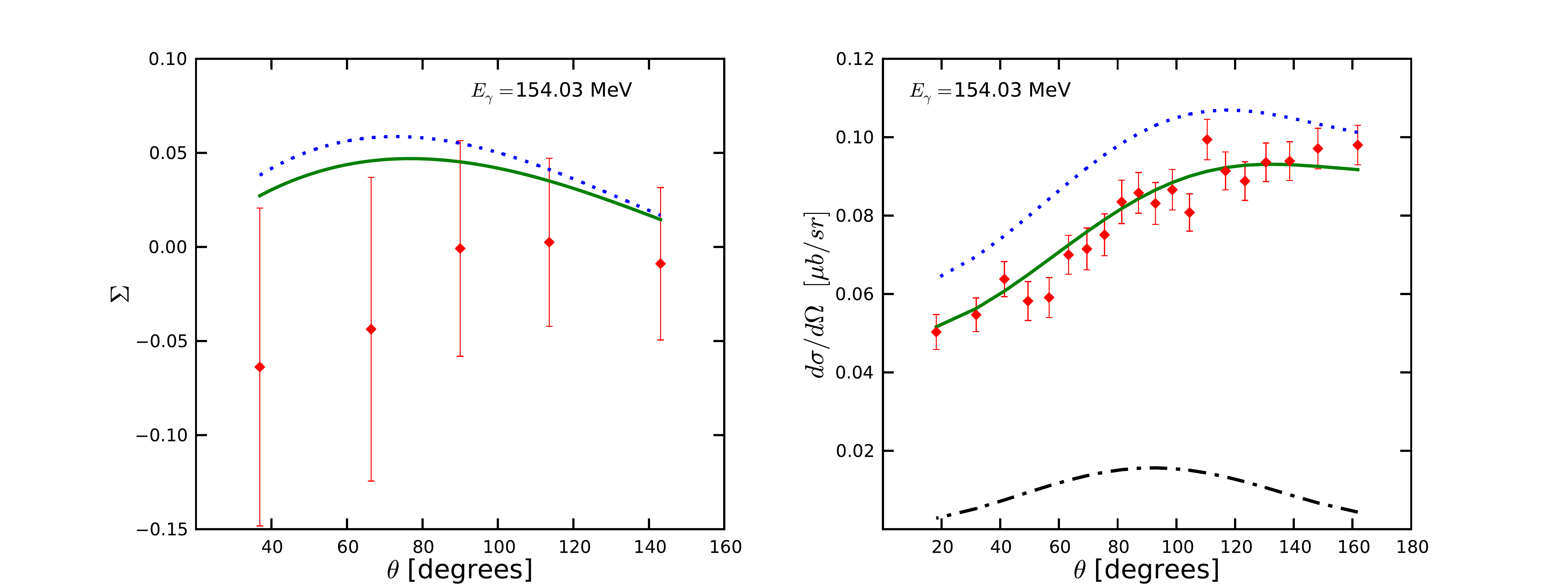}}\\
\subfigure{
\label{f8c}
\includegraphics[width=1.08\textwidth]{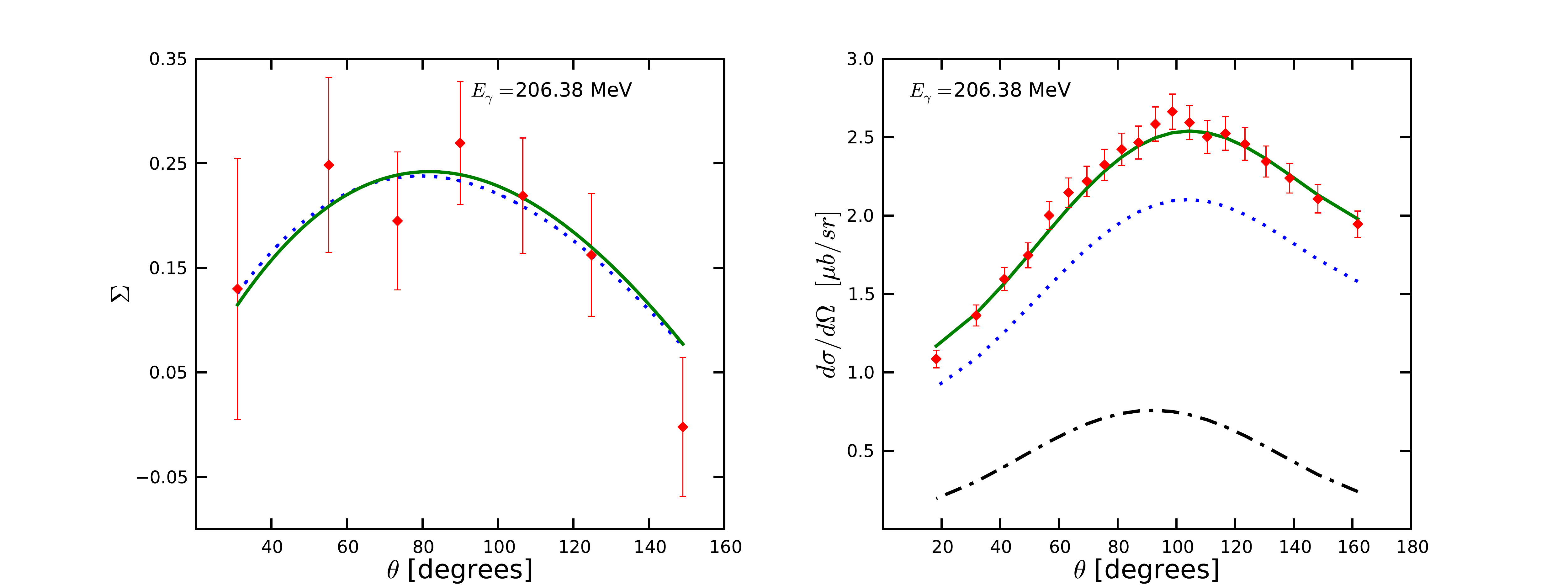}}
\end{center}
\caption{Photon asymmetry and differential cross section as a function of the pion angle. Solid line: full model, dotted line: full model without $\Delta$, dash-dotted line: only $\Delta$. Experimental points from Refs.~\cite{Hornidge:2012ca, Hornidge:pri}.}\label{fig:8}
\end{figure}

 As is obvious from the low $\chi^2$ value, the overall agreement of the full model is good.
In Fig.~\ref{fig:8}, we show two extreme (low and high energy) plots of the data set compared with our model with and without $\Delta$ mechanisms to better observe their effects.
Without $\Delta$, the model has a too slow energy dependence, which cannot reproduce the rapid increase of the cross section. Thus, the best fit occurs when the model overestimates the low energy data and underestimates the higher energy ones. On the other hand, the $\Delta$ mechanisms lead to a sharper slope allowing for a good fit over the whole energy range. We also show in the figure the cross sections obtained taking only the $\Delta$ contribution, which is relatively small at all energies. 

In summary, we have studied  the neutral pion photoproduction on the proton near threshold in covariant chiral perturbation theory with the explicit inclusion of $\Delta$ degrees of freedom at $O(p^3)$ and using the EOMS renormalization approach. We have compared our model with the recent and precise data from Ref.~\cite{Hornidge:2012ca} finding a good agreement for both the cross section and 
the linearly polarized photon asymmetry. We have also shown that the inclusion of the $\Delta$ resonance mechanisms substantially improves the agreement with data  over a wider energy  range than in previous calculations both in HB and covariant ChPT.

\begin{acknowledgments}
 This research was supported by the Spanish Ministerio de Econom\'ia y Competitividad and European FEDER funds under Contract No. FIS2011-28853-C02-01, by Generalitat Valenciana under Contract No. PROMETEO/20090090 and by the EU HadronPhysics3 project, Grant Agreement No. 283286. A.N. Hiller Blin acknowledges support from the Santiago Grisol\'ia program of the  Generalitat Valenciana. We thank D. Hornidge
for providing us with the full set of data from Ref.~\cite{Hornidge:2012ca}.

\end{acknowledgments}


\end{document}